\documentclass[runningheads]{llncs}

\usepackage{graphicx}
\usepackage{todonotes}
\usepackage{booktabs}
\usepackage{multirow}
\usepackage{makecell}
\usepackage[draft]{minted}
\usepackage{url}

\input{./_minted-ms/default.pygstyle}

\begin{document}
\title{Patapasco: A Python Framework for Cross-Language Information Retrieval Experiments}
\titlerunning{Patapasco: A Python Framework for CLIR Experiments}
\author{Cash Costello\inst{1}\orcidID{0000-0003-1003-4645} \and
Eugene Yang\inst{1}\orcidID{0000-0002-0051-1535} \and \\
Dawn Lawrie\inst{1}\orcidID{0000-0001-7347-7086}\and
James Mayfield\inst{1}\orcidID{0000-0003-3866-3013}}
\authorrunning{C. Costello et al.}
\institute{Human Language Technology Center of Excellence, Johns Hopkins University, Baltimore MD 21211, USA \\
\email{\{ccostel2,eugene.yang,lawrie,mayfield\}@jhu.edu}\\
\url{https://hltcoe.jhu.edu/}}
\maketitle              %
\begin{abstract}
While there are high-quality software frameworks for information retrieval experimentation,
they do not explicitly support cross-language information retrieval (CLIR).
To fill this gap, we have created Patapsco, a Python CLIR framework. This framework specifically
addresses the complexity that comes with running experiments in multiple languages.
Patapsco is designed to be extensible to many language pairs, to be scalable to large document collections,
and to support reproducible experiments driven by a configuration file.
We include Patapsco results on standard CLIR collections using multiple settings.

\end{abstract}
\section{Introduction}
The introduction of neural ranking methods to information retrieval (IR) research has led to the adoption of two-stage pipelines.
In the first stage, documents are retrieved from an inverted index and ranked according to a scoring function such as BM25.
Those documents are then re-ranked in a second stage using a slower neural model trained for the task.
Several software frameworks have been created around multi-stage ranking,
including Pyserini~\cite{lin2021pyserini}, PyTerrier~\cite{macdonald2020declarative}, and OpenNIR~\cite{macavaney2020opennir}.
These frameworks standardize the first stage,
allowing researchers to focus on developing better-performing neural ranking models.
They also support reproducibility and comparison of performance across different models with standardized data sets and settings.
These frameworks are designed with the assumption that the queries and documents are written in the same language.

Cross-language information retrieval (CLIR) is the retrieval of documents in one language based on a search query in another language. 
In this setting, the system needs to be aware of the language of the queries and the documents,
what types of processing are supported in those languages,
and how the language barrier is to be crossed.
Existing frameworks were not designed to handle these complexities.
However, with the advent of high-quality machine translation and the success of neural ranking methods for IR,
neural CLIR experimentation needs to be supported.
The Patapsco framework implements the modern retrieve-and-rerank approach for the CLIR use case by using
the existing Pyserini framework and extending it to support CLIR.
Patapsco was successfully used in the summer of 2021 at a CLIR workshop involving more than fifty participants. 
Such a large scale workshop of which many participants were new to 
CLIR research demonstrated how Patapsco enables sophisticated CLIR experiments and lowers barriers to entry for newcomers to the field

\section{System Overview}
Patapsco\footnote{Patapsco is available at \url{https://github.com/hltcoe/patapsco}.
A video demonstration is at \url{https://www.youtube.com/watch?v=jYj1GAbABBc}.} 
is built on top of Pyserini
and maintains Pyserini's design goal of reproducible experiments.
Patapsco adds extensive language-specific preprocessing, and scalability through parallel processing.
An experiment is described in a configuration file.
This file is used to generate the pipeline,
which begins with a standard 
information retrieval setup of a document collection, a topic file, and, if available, relevance judgments.
The pipeline is reproducible from that configuration file.
For CLIR support, Patapsco maintains the language metadata for documents and queries throughout the pipeline to enable language-specific processing.
This is a key feature needed to support CLIR experimentation 
that is not present in existing frameworks.

Patapsco handles conversion of topics into queries, ingest and normalization of documents and queries, inverted index construction,
retrieval of an initial results set, reranking, and scoring.
A pipeline can be run to completion to compute scores through pytrec\_eval~\cite{VanGysel2018pytreceval},
or it can be stopped early to create artifacts that support development and training.
A pipeline can also start where a previous pipeline stopped.
For example, a set of experiments may use the same preprocessing and indexing, and differ only in retrieval and reranking.

\subsection{Text Preprocessing}
As is standard in IR systems, Patapsco normalizes the documents and queries used in an experiment.
Character-level normalization, such as correcting Unicode encoding issues and standardizing diacritics, is applied first.
This is followed by token-level normalization,
which includes stop word removal and stemming.
Preprocessing is language-dependent and specified in the configuration file:

\input{./_minted-ms/57025AC160FC1CF36F4142C90D316594A98ABAE5F0548CD42DE6688C75F7E370.pygtex}
% \begin{minted}{yaml}
% documents:
%   input:
%     format: json
%     lang: rus
%     path: /data/clef02/rus_docs.jsonl
%   process:
%     normalize:
%       lowercase: true
%     tokenize: spacy
%     stopwords: lucene
%     stem: false
% \end{minted}

Patapsco supports both the rule-based tokenization and stemming used by most IR systems,
and neural models from toolkits like spaCy~\cite{spacy} and Stanza~\cite{qi2020stanza}.

\subsection{Retrieval}
Patapsco uses Pyserini to retrieve initial document results sets from the inverted index.
Users can select between BM25 and a query likelihood model using Dirichlet smoothing.
Query expansion using RM3 is also available.

Two approaches to CLIR are document translation and query translation (both of which need to be produced externally).
To support the latter use case, 
Patapsco's default JSONL topic format can hold multiple translations per query.
A third CLIR approach is to project the query into the language of the documents using a probabilistic translation table.
This method, called Probabilistic Structured Query (PSQ)~\cite{darwish2003probabilistic},
projects each term in the query to multiple target language terms with associated weights.
Patapsco implements PSQ as a extension on Lucene.

\subsection{Reranking}
Researchers working on neural reranking have the option of registering their reranker with Patapsco and having it run directly in the pipeline,
or implementing a command-line interface that is executed from the pipeline.
A primary advantage of the command line interface is the avoidance of any dependency conflicts with Patapsco,
which is a common issue with machine learning frameworks in Python.

The reranker is passed the query, the list of document identifiers, and access to a document database.
This database contains the original documents,
since the tokenization and normalization required by the word embedding is likely different than that used for building the inverted index.

\subsection{Parallel Processing}
To support large document collections with the added computation required by neural text processing models,
Patapsco includes two ways to use parallel processing:
multiprocessing and grid computing.
Each divides the documents into chunks that are processed in separate processes and then assembled in a map-reduce job.
Both slurm and qsub are supported for grid computing;
Patapsco manages the entire job submission process;
the user only has to select the queue and the number of jobs.

\section{Evaluation}

A set of baseline experiments conducted with Patapsco is presented in Table~\ref{tab:eval}, which reports mean average precision (MAP),
nDCG at rank 1000, and recall at rank 1000, where we observe the classic results of translating the documents yields better effectiveness than translating the queries. 
We evaluate the probabilistic structured query~\cite{darwish2003probabilistic} and translation approaches
on three widely-used CLIR collections:
CLEF Russian and Persian~\cite{peters2001european}, and NTCIR Chinese collections\footnote{\url{http://research.nii.ac.jp/ntcir/permission/ntcir-8/perm-en-ACLIA.html}}.
Each query is formed by concatenating the topic title and description. 
The experiments are executed on a research cluster with 20 parallel jobs during indexing and one job for retrieval.
The effectiveness is on par with the implementations from other studies. 
The running time ranges from several minutes to an hour,
depending on the size of the collection and the tokenization used in the experiment.
This fast running time enables large ablation studies.
The index is supported by Pyserini~\cite{lin2021pyserini}, which is a framework designed for large-scale IR experiments.
The memory footprint is consequently minimal, ranging from 2 to 3 GB in our experiments.

\begin{table}[t]
\caption{Baseline runs using Patapsco. QT/DT/HT stand for machine query translation, machine document translation, and human query translation. }\label{tab:eval}
\newcommand{\z}{\phantom{0}}
\setlength\tabcolsep{0.45em}
\centering
\begin{tabular}{l|ccc|ccc|ccc}
\toprule
& \multicolumn{3}{c|}{CLEF Persian}
& \multicolumn{3}{c|}{CLEF Russian}
& \multicolumn{3}{c}{NTCIR Chinese} \\
Model & MAP &  nDCG &  R@1k & MAP &  nDCG &  R@1k & MAP &  nDCG &  R@1k \\
\midrule
PSQ &  0.1370 &  0.3651 &  0.4832 &    0.2879 &  0.4393 &  0.7441 &   0.1964 &  0.3752 &  0.5867 \\
QT  &  0.2511 &  0.5340 &  0.6945 &    0.3857 &  0.5527 &  0.9268 &   0.2186 &  0.3953 &  0.6201 \\
DT  &  0.3420 &  0.6808 &  0.8501 &    0.3408 &  0.5151 &  0.8881 &   0.3413 &  0.5627 &  0.8110 \\
HT  &  0.4201 &  0.7476 &  0.9175 &    0.3975 &  0.5623 &  0.9401 &   0.4810 &  0.6840 &  0.9125 \\
\bottomrule
\end{tabular}
\vspace{-1em}
\end{table}

\section{Conclusion}
Patapsco brings recent advances in IR software frameworks and reproducibility to the CLIR research community.
It provides configuration-driven experiments, parallel processing for scalability, a flexible pipeline, and a solid baseline for performance evaluations.
Patapsco's configuration file fully documents each experiment, making them simple to reproduce. 
Patapsco enables sophisticated CLIR experiments and lowers barriers to entry for newcomers to the field.
During a CLIR workshop in the summer of 2021, researchers from outside of information retrieval successfully ran CLIR experiments
because of the ease of experimentation provided by Patapsco.
\bibliographystyle{splncs04}
\bibliography{ms}

\end{document}